\documentclass[secnumarabic,aps,prl,reprint,groupedaddress]{revtex4-1}

\usepackage[english]{babel}
\usepackage{graphicx}
\usepackage{color}
\usepackage{amsmath}
\usepackage{tabularx}

\begin{document}
\title{Brillouin Light Scattering of Spin Waves Inaccessible with Free-Space Light}

\author{Ryan Freeman$^1$}
\author{Robert Lemasters$^1$}
\author{Tomi Kalejaiye$^1$}
\author{Feng Wang$^1$}
\author{Guanxiong Chen$^1$}
\author{Jinjun Ding$^2$}
\author{Mingzhong Wu$^2$}
\author{Vladislav E. Demidov$^3$}
\author{Sergej O. Demokritov$^3$}
\author{Hayk Harutyunyan$^1$}
\author{Sergei Urazhdin$^1$}

\affiliation{$^1$Department of Physics, Emory University, Atlanta, GA, USA.}
\affiliation{$^2$Colorado State University, Fort Collins, USA}
\affiliation{$^3$University of M\"unster, M\"unster, Germany}

\begin{abstract}
	Micro-focus Brillouin light scattering is a powerful technique for the spectroscopic and spatial characterization of elementary excitations in materials. However, the small momentum of light limits the accessible excitations to the center of the Brillouin zone. Here, we utilize a metallic nanoantenna fabricated on the archetypal ferrimagnet yttrium iron garnet to demonstrate the possibility of Brillouin light scattering from large-wavevector, high-frequency spin wave excitations that are inaccessible with free-space light. The antenna facilitates sub-diffraction confinement of electromagnetic field, which enhances the local field intensity and generates momentum components significantly larger than those of free-space light. Our approach provides access to high frequency spin waves important for fast nanomagnetic devices, and can be generalized to other types of excitations and light scattering techniques.
\end{abstract}

\maketitle

Light-matter interaction provides an indispensable approach to studying physical, electronic, and magnetic properties of materials. In inelastic light scattering, exemplified by Brillouin light spectroscopy (BLS), the spectrum of scattered light is analyzed to determine the dispersion of elementary excitations such as phonons, excitons, or magnons (quanta of spin waves in magnetic materials), as well as their transport and relaxation~\cite{BLS_for_material_properties,BLS_for_material_properties2,BLS_for_material_properties3,BLS_for_material_properties4,BLS_magnons,Sermage1979,demokritov2001brillouin}. Micro-focus BLS ($\mu$-BLS), which utilizes a focused probing light spot~\cite{DemDem08, microBLSreview,Demidov2015,PhysRevLett.107.107204}, has provided an unprecedented ability to study magnetization dynamics at nanoscale~\cite{Demokritov2006, microBLSreview,DemDem08,demokritov2001brillouin}. $\mu$-BLS is sensitive exclusively to long-wavelength spin waves, which represents a significant obstacle to progress in the understanding of short-wavelength spin wave-modes that control fast magnetization dynamics at nanoscale~\cite{nonlinear_magnon_scattering}, which is important for the development of faster and more efficient magnetic nanodevices.

The spectroscopic limitations of $\mu$-BLS originate from the small momentum of visible light, which appears to be inherent to this technique. Here, we demonstrate that sub-diffraction confinement of electromagnetic field by a metallic nanoantenna fabricated on top of the magnetic film can efficiently overcome this limitation. We show experimentally, and with simulations, that the antenna generates large-momentum components of electromagnetic field not present in free-space light, providing information about the high-frequency states not accessible to the standard BLS. Our approach can be extended to other types of excitations such as phonons, and other inelastic scattering techniques such as Raman spectroscopy. 

\begin{figure}
	\includegraphics[width=\columnwidth]{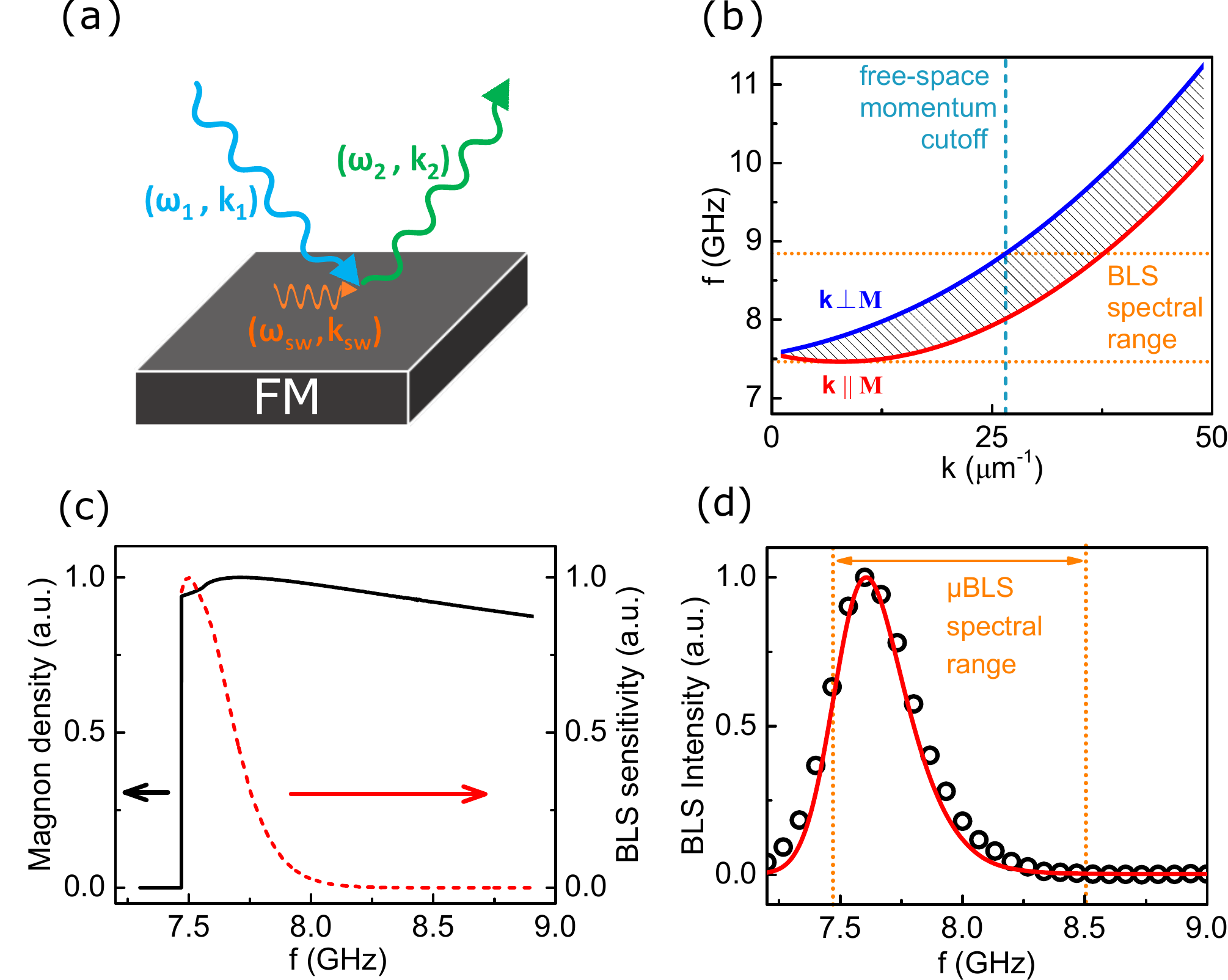}
	\caption{ (a) Schematic of BLS. (b) Spin-wave manifold for small wavenumbers, calculated using the analytical spin wave theory~\cite{SpinWaveTextbook}. The vertical dashed line indicates the largest wavevector accessible to BLS at the probing light wavelength $\lambda=473$~nm, due to momentum conservation. The horizontal lines show the accessible spectral range. (c) Calculated spectral sensitivity of diffraction-limited $\mu$-BLS (dashed curve) and the spectral density of thermal magnons in YIG(30) at $T=295$~K (solid curve). (d) Experimental (symbols) and calculated (curve) $\mu$-BLS spectrum of thermal spin waves in YIG(30),  at an in-plane filed $H=2$~kOe, and $T=295$~K. Dashed lines indicate the bottom of the spin-wave spectrum and the high-frequency BLS sensitivity cut-off.}\label{fig:BLS_limits}
\end{figure}

In an elementary scattering event underlying the anti-Stokes BLS in magnetic materials, a magnon is annihilated, transferring its energy and momentum to the scattered photon [Fig.~\ref{fig:BLS_limits}a]. This process satisfies energy and momentum conservation, expressed in terms of the frequencies and the wavevectors as $f_{2}=f_{1}+f_{sw}$, and $\vec{k}_{2}=\vec{k}_{1}+\vec{k}_{sw}$. Here, $f_{sw}$, $f_1$, and $f_2$ are the frequencies of the spin wave, the incident and the scattered photons, respectively, and $\vec{k}_{sw}$, $\vec{k}_{1}$, $\vec{k}_{2}$ are the corresponding wavevectors. Conservation laws impose stringent constraints on the quasiparticle states accessible to BLS. For the probing light wavelength $\lambda=473$~nm, the largest wavenumber of spin waves accessible to BLS, achieved in the backscattering geometry, is $k_{max}=4\pi/\lambda=26.6$~$\mu$m$^{-1}$ [dashed vertical line in Fig.~\ref{fig:BLS_limits}b], less than $0.1\%$ of the typical Brillouin zone size. The corresponding spectral range is limited to the states only very close to the bottom of the spin-wave spectrum. For example, in a 30nm thick film of an archetypical insulating ferrimagnet, yttrium iron garnet (YIG(30))~\cite{YIG_fab,YIG_Growth}, the accessible range, counted from the lowest spin wave frequency to the highest observable frequency, is only $1.38$~GHz at an in-plane field $H=2$~kOe [Fig.~\ref{fig:BLS_limits}b], even though the spectrum itself extends into the THz range~\cite{White2007,THZ_spinwave}.

\begin{figure}
	\includegraphics[width=\columnwidth]{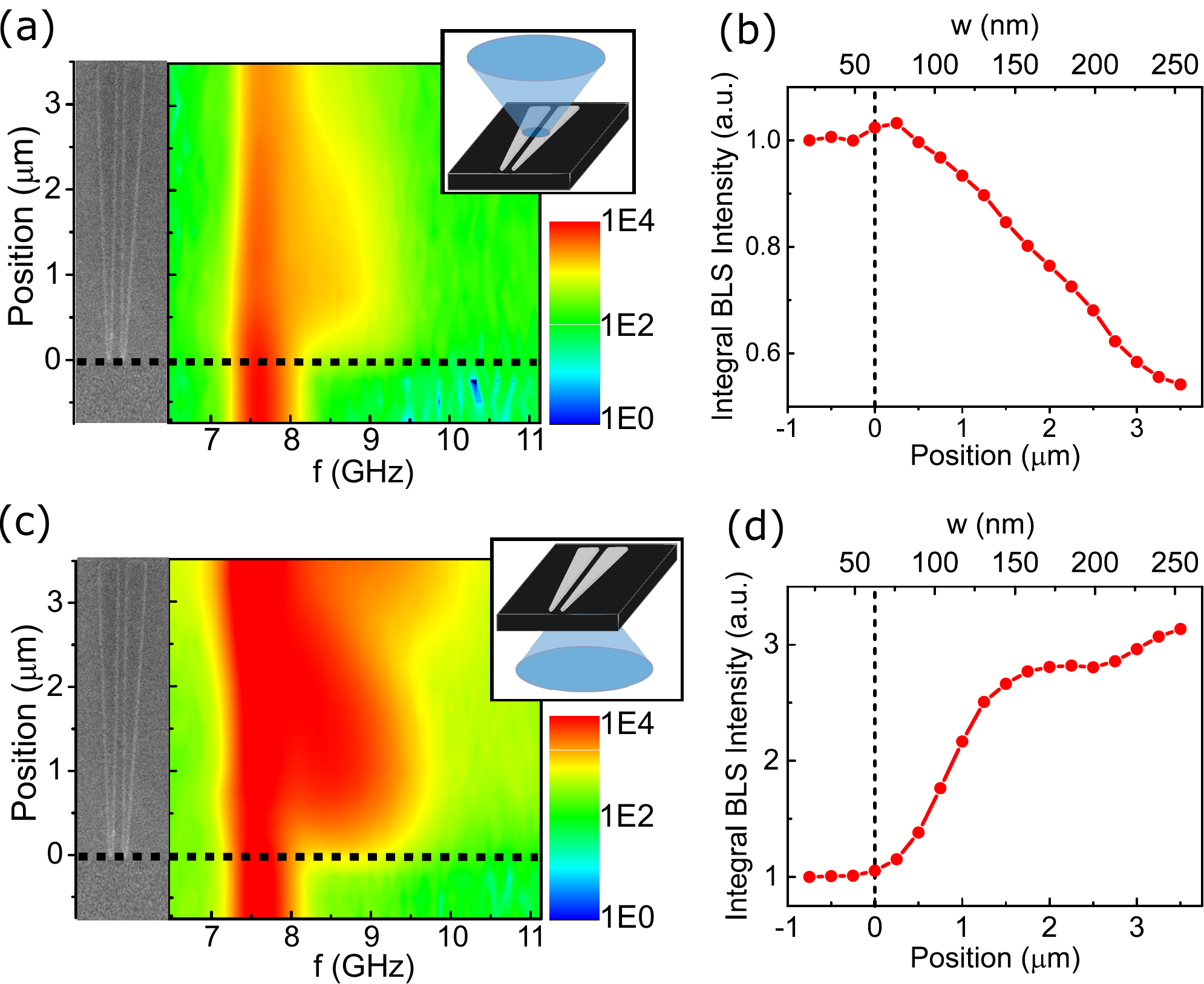}
	\caption{(a,c)  Pseudo-color maps of the $\mu$-BLS spectra of thermal spin waves at $295$~K, as a function of the position of the probing laser spot along the bi-wire antenna, with the beam incident from above (a) and from the substrate side (c). (b,d) Corresponding integral BLS intensity as a function of position along the antenna [bottom scale] and $w$ [top scale]. Insets in a,c, are schematics of the corresponding experimental layouts. In all measurements, the incident beam polarization and the in-plane field $H=2$~kOe were parallel to the wires.}\label{fig:BLS_exp}
\end{figure}

Here, we experimentally demonstrate that the spectroscopic limits of BLS can be overcome by utilizing the near-field redistribution of the electric field facilitated by a nanoscale metallic antenna fabricated on the magnetic system~\cite{supplementary}. We utilized an Al(30) biwire antenna fabricated on the YIG(30) film deposited on the GGG substrate\cite{supplementary}. The gap between the wires is $d=120$~nm, and the width $w$ of the wires was gradually varied from $90$~nm to $285$~nm along the antenna [SEM micrograph in Fig.~\ref{fig:BLS_exp}a, left], allowing us to probe the dependence of BLS signals on $w$ by scanning the probing spot along the antenna, to elucidate the mechanisms underlying the observed effects. We observed similar effects for a metallic ferromagnet Permalloy, confirming their generality~\cite{supplementary}.

To gain a qualitative insight into the inelastic light scattering in the presence of such a nanoantenna, we first approximate its effects as simple partial blocking of the incident light. Below, we present quantitative analysis that does not rely on such a simplifying assumption. We consider the diffraction-limited light spot with the half-width $\sigma$ incident on the film from above, and assume that $w\gg\sigma$, so that only a strip of width $d$ is not blocked. The resulting distribution of the momentum of the E-field is described by the Fraunhofer pattern, exhibiting maxima at $k=(2n+1)\pi/d$, with peak amplitude decreasing exponentially but remaining finite at any integer $n$~\cite{antennas}. Thus, the antenna generates large-momentum near-field components of electromagnetic field not present in the free-space light, which may overcome the restrictions on the spectral range of BLS illustrated in Fig.~\ref{fig:BLS_limits}.

We note that the effects of the antenna are not limited to light blocking. The interaction between the electric field of light and electrons in the antenna is expected to result in both the concentration of the electromagnetic energy of the light into the antenna's near-field and the generation of high-momentum components of field due to the spatial modulation of the field in the vicinity of the antenna~\cite{pohl2012near,Effective_Wavelength,amendola2017surface,MOPlasmonics1,MagnetoPlasmonics_Review,Original_SEBLS_Experiment,Utegulov_SEBLS}. These effects are confirmed by the data and analysis presented below.

Figure~\ref{fig:BLS_exp}a shows a pseudo-color map of the $\mu$-BLS spectra acquired with the probing spot incident from the free-space side of the structure. The main feature of these data is the abrupt broadening of the spectra, once the spot becomes positioned above the antenna, whose edge is indicated by the dashed line. The intensity decreases with increasing $w$, as expected due to the partial blocking of the beam by the nanowires. These trends are reflected by the analysis of the total BLS intensity, which exhibits  an overall decrease with increasing nanowire width as the probing spot is scanned along the antenna [Figure~\ref{fig:BLS_exp}b].

The intensity enhancement and the spectral broadening are significantly more pronounced for illumination from the substrate side, Fig.~\ref{fig:BLS_exp}c. In addition to the primary peak observed without antenna, a large secondary peak appears at frequencies around $9$~GHz as the probing spot crosses the edge of the antenna. In contrast to illumination from above, the intensity of the primary peak does not decrease with increasing $w$. Meanwhile, the intensity of the secondary peak first increases, and then starts to decrease, and a third peak emerges in the spectrum at higher frequencies. The additional peaks lie outside the spectral range accessible to free-space $\mu$-BLS [see Fig.~\ref{fig:BLS_limits}d], demonstrating that the nanowire antenna generates large-momentum components of optical field not present in free-space light.

Figure~\ref{fig:BLS_exp}c shows that the onset of large additional peaks in the BLS spectrum is not accompanied by the reduction of the BLS intensity for the illumination from the substrate side. On the contrary, the total BLS intensity increases by more than a factor of $3$ compared to the free-space illumination as the nanowire width $w$ approaches $300$~nm, Fig.~\ref{fig:BLS_exp}d. A pronounced bump in the intensity around $w=150$~nm is associated entirely with the onset of the secondary peak, demonstrating that the large-momentum components of optical field generated by the antenna are significant.

\begin{figure}
	\includegraphics[width=\columnwidth]{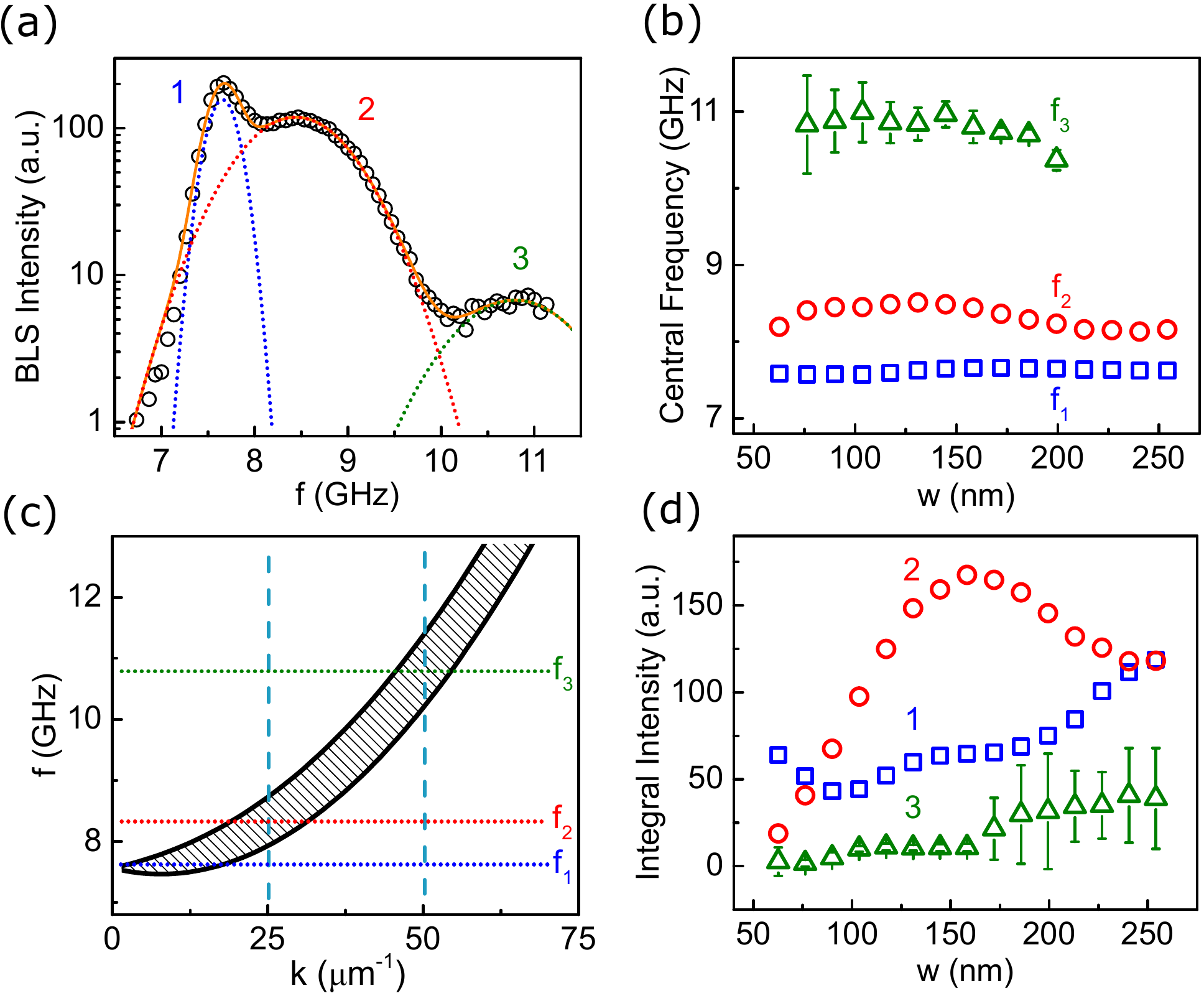}
	\caption{(a) $\mu$-BLS spectrum of thermal spin waves for $w=160$~nm (symbols), and its fitting with the sum of three Gaussian functions $g_i(f)=\frac{I_i}{\sqrt{2\pi}\Delta f_i}e^{-(f-f_i)^2/2\Delta f_i^2}$ (curve). The individual Gaussian curves are shown with dotted curves. (b,d) Central frequencies $f_i$ (b) and integral intensities $I_i$ (d) of the Gaussians vs. $w$. Where omitted, the error bars are smaller than symbols sizes. The peaks are labeled 1,2, and 3, as indicated in (a). The frequency of peak 3 is not shown for $w>200$~nm, because of the large fitting uncertainty. (c) Spin wave manifold superimposed with the central frequencies of the Gaussian peaks (horizontal lines), for $w=160$~nm. Vertical lines are guides for the eye. }\label{fig:BLS_analysis}
\end{figure}

We now quantitatively analyze the BLS spectra obtained with illumination from the substrate side. These spectra are characterized by three spectral features, labeled ``1'', ``2'', and ``3'' in Fig.~\ref{fig:BLS_analysis}a showing the spectrum for $w=160$~nm. We obtained a good quantitative fitting of these spectra by a sum of three Gaussian functions, as shown in Fig.~\ref{fig:BLS_analysis}a by a solid curve. The central frequency $f_1$ of the primary peak is independent of $w$, while the frequencies of the other peaks are approximately constant or exhibit a modest increase at small $w$, and then start decreasing at larger $w$ [Fig.~\ref{fig:BLS_analysis}b].

We establish the relationship between the observed spectral features and the momentum components of E-field generated by the antenna, by analyzing the spectral positions of the peaks relative to the spin-wave manifold, as shown in [Fig.~\ref{fig:BLS_analysis}c] for $w=160$~nm. As expected, the frequency $f_1$ of the first peak corresponds to the bottom of the manifold. Meanwhile, the central frequency $f_2=8.3$~GHz of the second peak corresponds to wavevectors around $k_{sw}\approx 25$~$\mu$m$^{-1}$, and the frequency $f_3$ of the third peak corresponds to the wavevectors around $k_{sw}\approx 50$~$\mu$m$^{-1}$, both outside the range accessible to $\mu BLS$ with free-space light. These results unambiguously demonstrate that the antenna enables one to overcome the fundamental spectroscopic limitations of free-space light scattering.

The efficiency of large momentum generation by the antenna is demonstrated by the analysis of the integral intensity of the peaks, Fig.~\ref{fig:BLS_analysis}d. The intensity of the primary peak initially slightly decreases, but then starts to gradually increase  with increasing nanowire width $w$, as expected from the overall increase of the incident light intensity due to the reflection from the antenna. In contrast, the intensity of the second peak rapidly increases, becoming three times larger than the primary peak near $w=160$~nm, and then starts to decrease at larger $w$. 

We note that the large intensity of the second peak is concealed in the spectra of Fig.~\ref{fig:BLS_analysis}c by its large width, which is over three times larger than the width of the primary peak~\cite{supplementary}. The strong nonmonotonic dependence of the second peak's intensity on $w$ indicates that constructive interference from different regions defined by the antenna plays a significant role in the generation of large momentum of E-field.

\begin{figure}
	\includegraphics[width=\columnwidth]{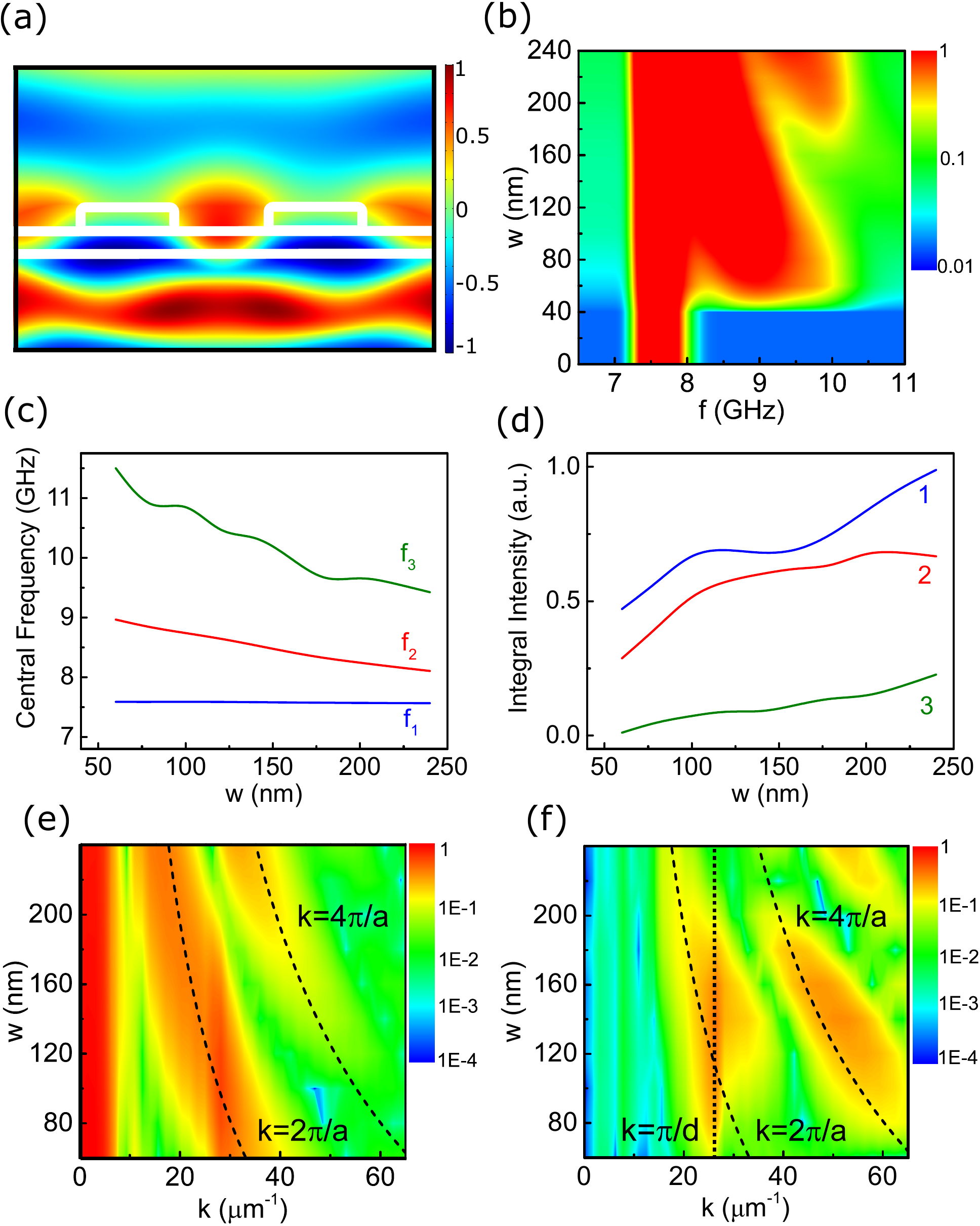}
	\caption{(a) Crossection of the spatial distribution of the component of the incident optical E-field directed along the antenna, for $w=d=120$~nm, and the polarization along the wire length. The boundaries of the antenna and the interfaces of the magnetic film are indicated by white lines. (b) Pseudo-color plot of the calculated $\mu$-BLS spectrum of thermal spin waves at $295$~K, as a function of $w$, for $d=120$~nm. (c,d) Dependencies on the wire width of the central frequencies (c) and integral intensities (d) of the peaks obtained by fitting the spectra in panel (b) with a sum of three Gaussians. (e) Spatial Fourier spectra of the component of the incident E-field along the wire, for polarization along the wire, vs $w$. (f) Spatial Fourier spectra of the out-of-plane component of the electric field, for polarization perpendicular to the wires. Dashed curves in (e), (f) show $k=2\pi/a$ and $4\pi/a$, as labeled, where $a=w+d$. The dotted vertical line in (f) shows $k=\pi/d$ }\label{fig:calculations}
\end{figure}

The mechanisms of the large momentum generation by the antenna are elucidated by the calculations of E-field distribution in the magnetic film, performed using the Comsol Multiphysics software. Figure~\ref{fig:calculations}a shows the calculated distribution of the local E-field of incident light in the cross-section of the antenna, for $w=120$~nm. This distribution is qualitatively distinct from the simple partial blocking by the antenna. The most prominent features are the narrow high-intensity spot in the center of the gap and two side-lobes localized under the wires characterized by the local polarization opposite to that of the central spot, with the phase of E-field reversing again outside the antenna. Thus, nanoantenna produces a rapid variation of E-field, which reverses its direction four times when crossing from one side of the antenna to the opposite side. The periodicity of the fundamental harmonic of this distribution is approximately $2\pi/(w+d)$, defining the dominant large-momentum component of E-field generated by the antenna.

To analyze the relation between the effects of antenna on E-field distribution and the $\mu$BLS spectra, we introduce a framework for the calculation of $\mu$BLS spectra. To the best of our knowledge, there is presently no accepted quantitative model of $\mu$BLS. Thus, calculations presented below not only provide insight into the effects of the nanoantenna, but also validate our model.

BLS can be analyzed in terms of the light scattering by the spin wave-induced dynamical modulations of the optical properties of the magnetic material, described by the time-dependent off-diagonal components of the permittivity tensor~\cite{wettling1975relation,landau1963course,SpinWaveTextbook,supplementary}. A focused laser beam produces an oscillating electric polarization  

\begin{equation}
\Delta \vec{P}(\vec{k}_2, \omega_2)=\sum_{\vec{k}_1} \Delta\hat{\epsilon}(\vec{k}_{sw},\omega_{sw} )\vec{E}(\vec{k}_1,\omega_1),
\end{equation}

where $\vec{k}_1=\vec{k}_2-\vec{k}_{sw}$ and $\vec{\omega}_1=\omega_2-\omega_{sw}$ reflect momentum and energy conservation, respectively, the summation is performed in the plane of the magnetic film in the k-space, $\Delta\hat{\epsilon}(\vec{k}_{sw},\omega_{sw} )$ is the permittivity tensor modulated by the spin waves, and $\vec{E}(\vec{k}_1,\omega_1)$ is the wavevector distribution of the incident field.

Without the nanoantenna, the amplitude of the scattered electromagnetic field with the 3d wavevector $\vec{k}_2$ characterized by the normal component $k_{2,y}=\sqrt{2\pi/\lambda_2-k_{2,xz}^2}$ is determined by the dispersion of light, re-radiated into the far-field due to the oscillating electric polarization $\Delta\vec{P}(\vec{k}_2)$. The nanoantenna changes the relationship between the local field in the magnetic film, $\vec{E}_{local,xy}$, and the far-field detected by the spectrometer, $\vec{E}_{far-field}$, which is determined by the antenna geometry and dielectric properties. These  effects can be described by a tensor $\hat{T}(\vec{k}_{local},\vec{k}_{far-field})$, 
such that $\vec{E}_{far-field}(\vec{k}_{far-field})=\sum_{k_{local}}\hat{T}(\vec{k}_{local},\vec{k}_{far-field})\vec{P}(\vec{k}_{local})$. Since the spin-waves modulate the off-diagonal components of the mageto-optic tensor, an analyzer with polarization  orthogonal to the incident light is used in the detection optics. Thus, the far-field detection sensitivity can be  described by a vector function $\vec{A}(\vec{k})$, with the direction orthogonal to the incident polarization. We combine the effects of the antenna and the detection optics into the detection sensitivity function 

\begin{equation}
\begin{split}
\vec{S}_j((\vec{k}_{local})=\\
\sum_{\vec{k}_{far-field},i}\vec{A}_i(\vec{k}_{far-field})\hat{T}_{ij}(\vec{k}_{local},\vec{k}_{far-field}),
\end{split}
\end{equation}

which is related to the electric field at the location of the optical detector via

\begin{equation}\label{E_detector}
E_{detector}\propto\sum_{\vec{k}_{local}}\vec{S}(k_{local})\cdot P_{local}(\vec{k}_{local}).
\end{equation}
This relationship illustrates the fact that, while the amplitude of the electric field incident on the detector is determined by multiple optical processes that include momentum redistribution by the antenna, and collection/focusing by the lens, the detected amplitude is simply proportional to the amplitude of the local electric field linearly scaled by these processes, with a wavevector-dependent scaling coefficient $\vec{S}$. This sensitivity function can be well approximated by the distribution $E_{local}(\vec{k})$ for the incident beam with the far-field polarization orthogonal to the incident beam~\cite{supplementary}.

The intensity of light at frequency ${\omega_1+\omega_{sw}}$, scattered by the spin-wave mode with frequency $\omega_{sw}$  is given by the absolute value squared of Eq.~(\ref{E_detector}). Summing over the incoherent contributions from different thermally excited spin wave modes, and taking into account the finite spectral resolution $R(f_{sw}(k_{sw}))=e^{-[f-f_{sw}(\vec{k}_{sw})]^2)/4\sigma^2}$of the BLS detection, we obtain the thermal $\mu$BLS intensity

\begin{widetext} \begin{equation}\label{eq:BLS_int}
	I_{BLS}(\omega)\propto\int d^2k_{sw}|\int d^2k_1 \vec{E}_1(\vec{k}_1)\delta\hat{\epsilon}_{12}(\vec{k}_{sw})\vec{S}(\vec{k}_1+\vec{k}_{sw})R(f_{sw}(k_{sw}))|^2,
	\end{equation}
\end{widetext}

Figure~\ref{fig:calculations}b shows the calculated dependence of the $\mu$-BLS spectra on the wire width, obtained using Eq.~(\ref{eq:BLS_int}) and the simulated E-field distributions such as that shown in Fig.~\ref{fig:calculations}a. These results capture the salient features of the data: A broad secondary peak that appears around $9$~GHz, and a smaller third peak that emerges at larger $w$. We obtained good fittings of the calculated spectra by the sum of three Gaussian functions. The frequency of the primary peak is independent of $w$, in agreement with the experimental data [Fig.~\ref{fig:BLS_analysis}b]. The frequencies of the secondary peaks are in overall agreement with the data, but their calculated dependence on $w$ is considerably stronger than observed. While the calculations underestimate the observed dramatic increase of the second peak's intensity, the calculated intensity of the primary peak increases by almost a factor of two when $w$ is increased from $90$~nm to $240$~nm [Fig.~\ref{fig:calculations}d], in a remarkable agreement with the data [Fig.~\ref{fig:BLS_analysis}d].

Analysis of the momentum distribution of E-fields provides insight into the effects of the antenna geometry on the generation of large-momentum components. Figure~\ref{fig:calculations}e shows the momentum distribution of the in-plane component of E-field for polarization parallel to the wires, while Fig.~\ref{fig:calculations}f shows the momentum distribution of the normal-to-plane component, for polarization perpendicular to the wires.  The former describes one of the contributions to the incident field term in Eq.~(\ref{eq:BLS_int}), while the latter - to the detection sensitivity function. These momentum distributions features a prominent secondary peak concentrated around $k=2\pi/(w+d)$, and a less pronounced peak around $k=4\pi/(w+d)$, consistent with our qualitative analysis of the spatial distribution in Fig.~\ref{fig:calculations}a. In addition, both distributions clearly exhibit enhanced amplitude at $k=\pi/d$, as shown by the dotted vertical line in Fig.~\ref{fig:calculations}f. Such enhancement is expected from the partial light blocking by a single slit of width $d$. The resulting Fraunhofer pattern of momentum distribution exhibits maxima around $k=(2n+1)\pi/d$, with integer $n$~\cite{antennas}, consistent with the observed enhancement. Thus, the secondary spectral peaks observed both in our experiment [see Fig.~\ref{fig:BLS_analysis}] and in calculations can be identified with the interplay between the two contributions discussed above generating momentum components $k=2\pi/(w+d)$ and $k=\pi/d$, which become mutually enhanced close to $w=d$. This effect can be alternatively described as a geometric enhancement due to the constructive interference of contributions from different regions defined by the antenna.

In summary, we have experimentally demonstrated and modeled inelastic light scattering by spin-wave excitations with large wavevectors, inaccessible to free-space optical techniques due to momentum conservation. Our approach utilizes a nanoscale metallic antenna to generate large-momentum components of optical field, and simultaneously increase its local intensity due to a combination of enhanced reflection and constructive interference. The spectral limitations associated with the small momentum of light, as well as the presented approach to large-momentum generation, are quite general. Thus, we expect that our approach can be extended to the optical studies of other types of excitations, such as phonons, excitons, and electronic transitions. The latter can facilitate more efficient optical detectors and photovoltaics based on indirect-bandgap semiconductors, where large momentum of electron-hole pairs results in an optical absorption bottleneck.  

\begin{acknowledgements}
We acknowledge support from the NSF ECCS-1804198, EFMA-1641989 and ECCS‐1915849 and from Deutsche Forschungsgemeinschaft (Project No. 416727653). H.H. and R.L. acknowledge funding from Department of Energy (DE-SC0020101).
\end{acknowledgements}

\bibliography{main_bib}
\bibliographystyle{apsrev4-1}

\end{document}